\documentclass[onecolumn,conference]{IEEEtran}

\usepackage{cite}

\usepackage{subfigure}  
\usepackage[pdftex]{graphicx}   

\usepackage[cmex10]{amsmath}
\newcommand{\eqs}{\begin{eqnarray*}}
\newcommand{\eqf}{\end{eqnarray*}}
\newcommand{\lef}{\left(}
\newcommand{\rig}{\right)}
\newcommand{\mas}{\begin{array}}
\newcommand{\maf}{\end{array}}
\newcommand{\deriv}{\textnormal{d}}

\usepackage{fancyhdr}
\fancypagestyle{firststyle}
{
  \fancyhf{}

  \fancyfoot[C]{978-0-692-51523-5
  }
}

\begin{document}
\title{Analytic solution of Ando's surface roughness model with finite domain distribution functions}

\author{\IEEEauthorblockN{Kristof Moors\IEEEauthorrefmark{1}\IEEEauthorrefmark{2},
Bart Sor\'ee\IEEEauthorrefmark{1}\IEEEauthorrefmark{3}\IEEEauthorrefmark{4} and
Wim Magnus\IEEEauthorrefmark{1}\IEEEauthorrefmark{3}}
\IEEEauthorblockA{\IEEEauthorrefmark{1}imec, Kapeldreef 75, B-3001 Leuven, Belgium}
\IEEEauthorblockA{\IEEEauthorrefmark{2}Institute for Theoretical Physics, KU Leuven, Celestijnenlaan 200D, B-3001 Leuven, Belgium}
\IEEEauthorblockA{\IEEEauthorrefmark{3}Physics Department, University of Antwerp, Groenenborgerlaan 171, B-2020 Antwerpen, Belgium}
\IEEEauthorblockA{\IEEEauthorrefmark{4}Electrical Engineering (ESAT) Department, KU Leuven, Kasteelpark Arenberg 10, B-3001 Leuven, Belgium\\Email: kristof@itf.fys.kuleuven.be}}

\maketitle
\thispagestyle{firststyle}

\begin{abstract}
Ando's surface roughness model is applied to metallic nanowires and extended beyond small roughness size and infinite barrier limit approximations for the wavefunction overlaps, such as the Prange-Nee approximation. Accurate and fast simulations can still be performed without invoking these overlap approximations by averaging over roughness profiles using finite domain distribution functions to obtain an analytic solution for the scattering rates. The simulations indicate that overlap approximations, while predicting a resistivity that agrees more or less with our novel approach, poorly estimate the underlying scattering rates. All methods show that a momentum gap between left- and right-moving electrons at the Fermi level, surpassing a critical momentum gap, gives rise to a substantial decrease in resistivity.
\end{abstract}

\section{Introduction}
Surface roughness (SR) is important in metallic nanowires, serving as interconnects, as a source of scattering and it is challenging to treat its effect on electron transport rigorously. The use of phenomenological parameters, such as the specularity parameter in the Fuchs-Sondheimer model \cite{fuchs1938conductivity,sondheimer1952mean}, or a classical description of electrons and wire boundaries are common practice. For larger diameters, these models predict a 1/width proportionality for the resistivity and this is also confirmed by experimental measurements \cite{josell2009size}. Because of the shortcomings of these models and lack of experimental data however, the resistivity scaling behavior and the impact of SR on the sub-10~nm scale is unknown. Especially in this regime a better model would be useful to understand and predict the effect of SR on electron transport, which could also help to interpret resistivity measurements and possibly extract the contribution from SR to the total resistivity.

Ando's SR model provides a way of treating electrons undergoing SR scattering quantum mechanically through Fermi's golden rule to obtain the electron scattering rates \cite{ando1982electronic}. The detailed properties of the roughness profile can be taken into account by statistically averaging the scattering rates over different profiles, specifying a SR standard deviation, correlation length and autocorrelation function. We have extended Ando's model beyond the Prange-Nee approximation \cite{prange1968quantum} and adapted it for metallic nanowires with realistic roughness sizes, while it was originally developed for a 2D electron gas of an inversion layer at a semiconductor-insulator interface. Even with these modifications, we could still find an analytic solution of the averaged SR matrix elements by making use of SR distribution functions on a finite domain. This allows us to perform accurate and fast metallic nanowire simulations with a wide range of roughness profile parameters \cite{moors2015surfaceroughness} and study the transport properties by using the scattering rates in a self-consistent multi-subband Boltzmann transport solver \cite{moors2014resistivity}.

\section{Ando's surface roughness model}
\begin{figure}[tb]
\centering
\subfigure[\ Subbands: $D\approx 2.15$~nm]{\includegraphics[width=0.4\linewidth]{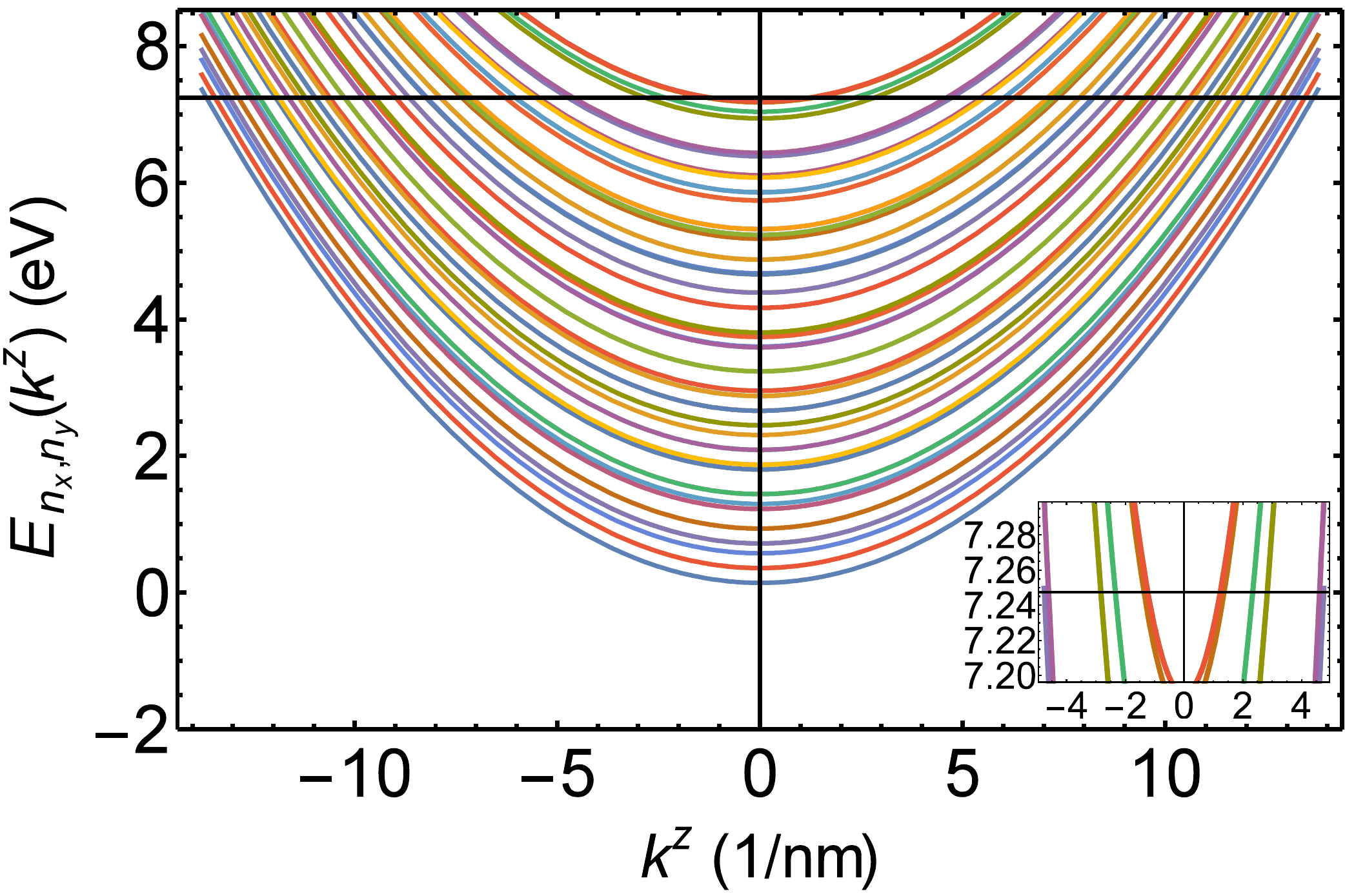}}
\subfigure[\ Subbands: $D\approx 3.25$~nm]{\includegraphics[width=0.4\linewidth]{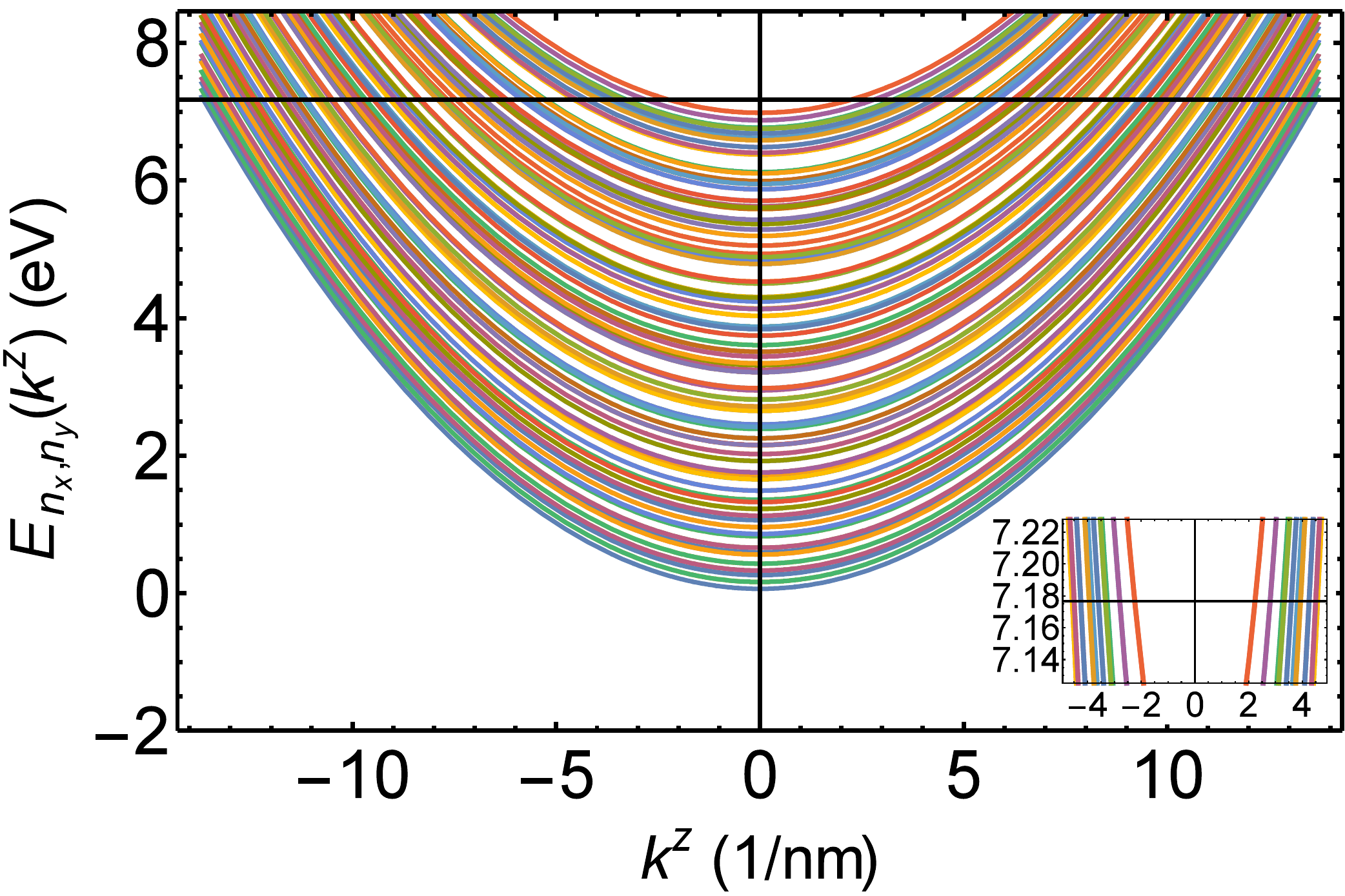}}
\caption{The subbands with states below the Fermi level of (a) $D\approx 2.15$~nm (b) $D\approx 3.25$~nm nanowire with square cross section ($L_x=L_y\equiv D$) are shown with a closer view around wavevector $k^z=0$ in the lower right corner. An effective mass equal to the free electron mass is considered and the Fermi level is obtained self-consistently with the conduction electron density of Cu: $n_e \approx 8.469\cdot 10^{28}$~m${}^{-3}$.}
\label{figSubbands}
\end{figure}

Before the SR matrix elements can be computed, the eigenstates in the ideal nanowire have to be obtained. We consider single-electron states with an isotropic effective mass ($m_e$) Hamiltonian in a 3D finite potential well (barrier height $U$) with rectangular cross section (sides $L_x, L_y$) and translational invariance along the transport direction (length $L_z \gg L_x, L_y$):
\begin{align*}
H_0 \lef \mathbf{r} \rig &=
\left\{
\begin{matrix}
\; - \dfrac{\hbar^2}{2 m_e} \nabla^2, \; \textnormal{if } 0 \leq x/y \leq L_{x/y} \; \& \; |z| \leq L_z/2 \\
\mkern-20mu - \dfrac{\hbar^2}{2 m_e} \nabla^2 + U, \; \textnormal{else} \qquad \quad \qquad \qquad \qquad
\end{matrix} \right. .
\end{align*}
The subbands are filled up to the electron density for the metal under consideration to obtain the Fermi level and this typically leads to a wide energy range (several eV) between the bottom of the lowest subband and the Fermi level; hence there are many possible state pairs for which to compute SR matrix elements (see Fig.~\ref{figSubbands}). For close to room temperature conditions and SR scattering, being well described as an elastic scattering process, a zero-temperature approximation works very well. In this way we restrict the possible states to the finite number of states crossing the Fermi level, two per parabolic subband.

\begin{figure}[tb]
\centering
\includegraphics[width=0.425\linewidth]{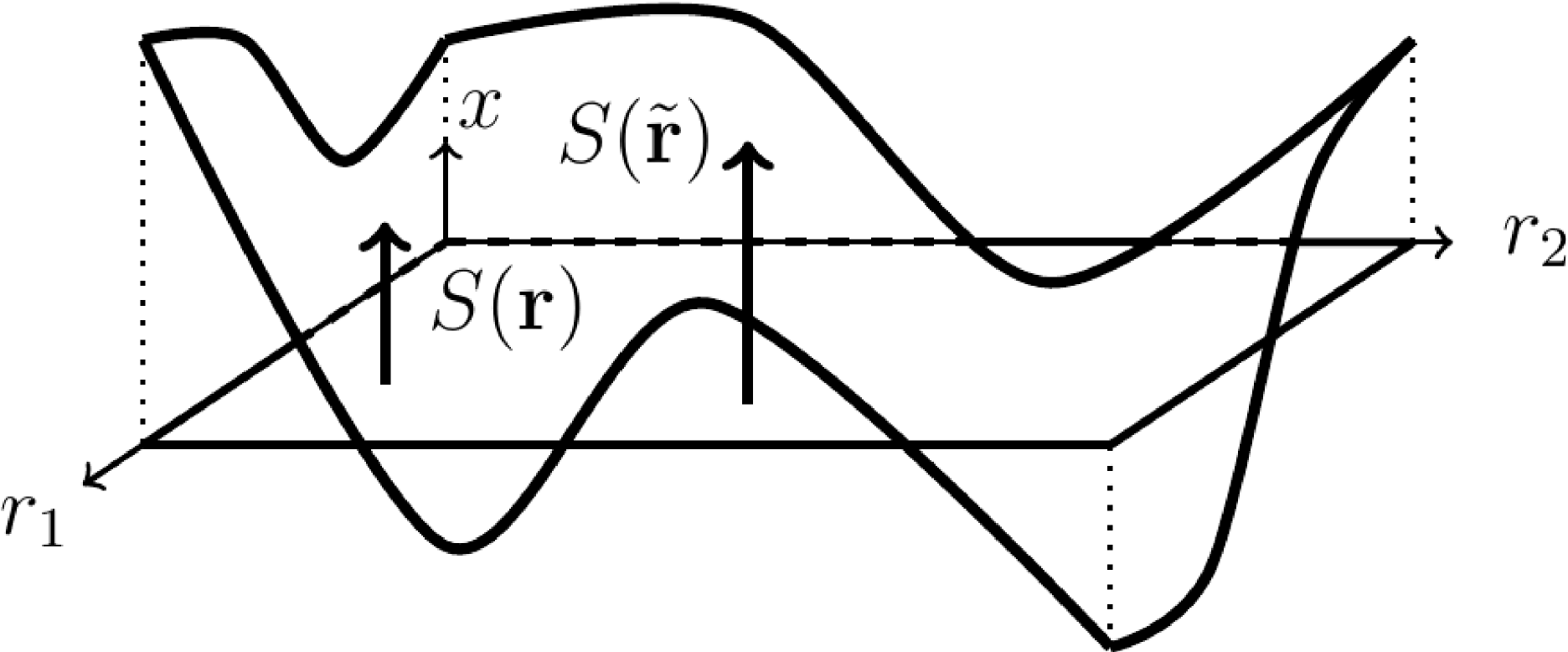}
\caption{The rough surface is described by a surface function $S$ that gives the height w.r.t. the smooth boundary surface at every point $\mathbf{r}$ on that surface. This function is never used explicitly to obtain the scattering rates, only its statistical properties.}
\label{figSR}
\end{figure}

\begin{figure}[tb]
\centering
\subfigure[\ Rough boundary position: $x=-0.5$~nm]{\includegraphics[width=0.45\linewidth]{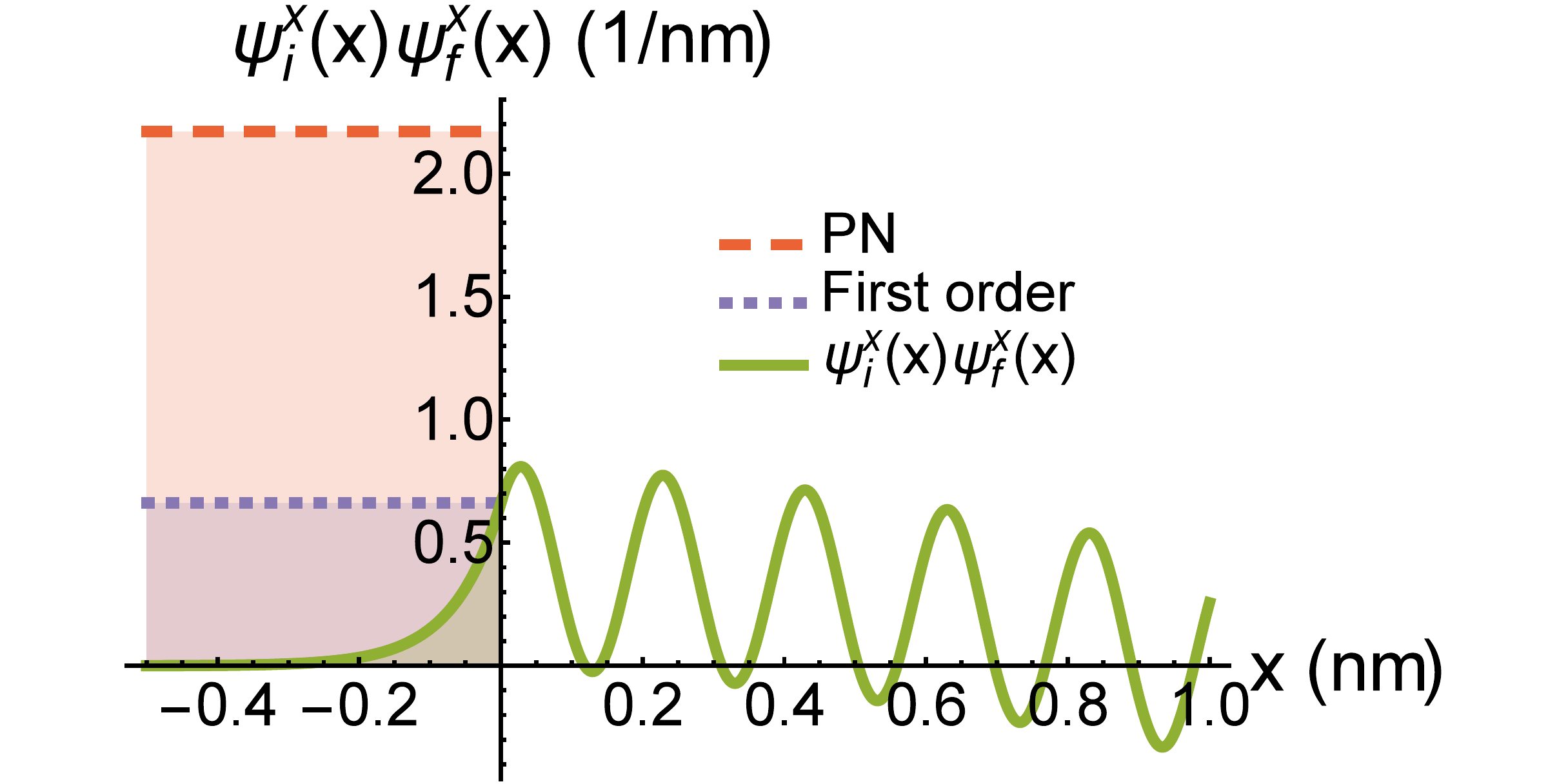}}
\subfigure[\ Rough boundary position: $x=1$~nm]{\includegraphics[width=0.45\linewidth]{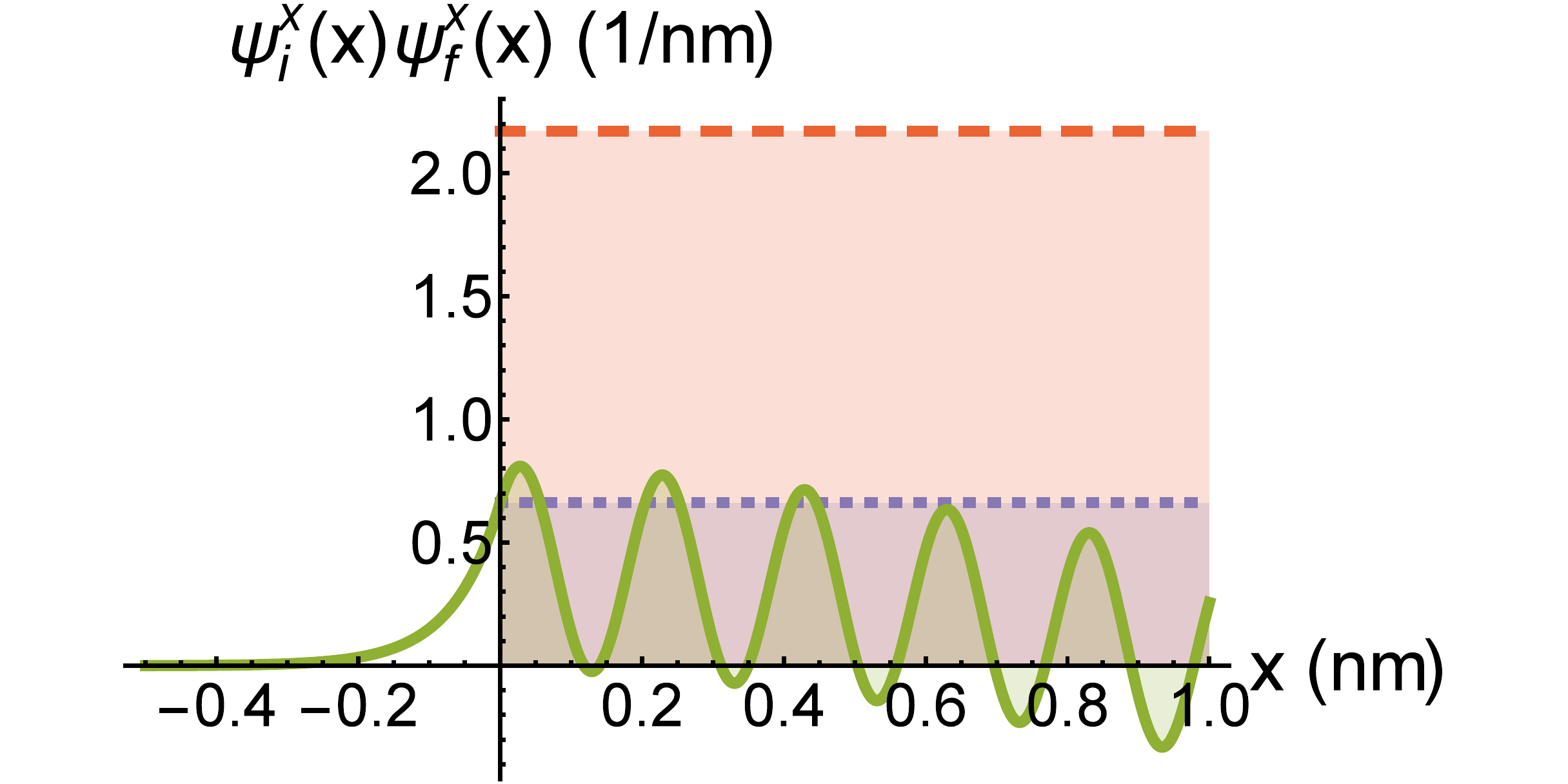}}
\caption{A product of wave functions for a wire $L_x=L_y \approx 2.15$~nm and the integral up to the rough boundary position are shown as a shaded area under the curve for with $i$ and $f$ states at the Fermi level with highest and next to highest $k^x$ value. The area under the curve of the wave function product should enter the SR matrix element of Ando's model, but the first order and Prange-Nee (PN) approximation replace this area by a rectangular box with height given by the value of the wave function product at $x=0$ or its infinite barrier limit respectively.}
\label{WFInt}
\end{figure}

Ando's SR model can be derived by treating the difference in potentials between a smooth and a rough nanowire as the perturbation potential ($V^\textnormal{\tiny SR}$) in Fermi's golden rule. We consider an example of a rough surface, described by the surface function $S(\mathbf{r})$ (see Fig.~\ref{figSR}) with $x$ denoting the direction normal to the flat surface, which leads to:
\begin{align*}
V^\textnormal{\tiny SR}(x,\mathbf{r}) &= H_0 \lef x - S(\mathbf{r}), \mathbf{r} \rig - H_0 \lef x, \mathbf{r} \rig.
\end{align*}
The potential is then easily generalized to SR for every nanowire boundary surface, although the effects of corners where two boundaries meet are neglected. The matrix elements have to be computed with wave functions of the form $\psi^x(x) \psi^y(y) e^{ik^z z}/\sqrt{L_z}$, with $k^z$ the wavevector along the transport direction and $\psi^{x}(x) = A \sin\lef k^{x} x \rig$ the wave function along a confinement direction when $x$ is inside the potential well and the sine replaced with an exponential tail when outside. The matrix elements $\langle i \mid V^\textnormal{SR} \mid f \rangle$ that enter Fermi's golden rule are functionals of $S(\mathbf{r})$ with a highly nonlinear dependence because of the convolution of $V^\textnormal{\tiny SR}$ with the oscillating wave functions. The matrix element for SR of the $x=0$ surface is given by:
\begin{align*}
\langle i \mid V^{\textnormal{\tiny SR}}_{(x=0)} \mid f \rangle &= \frac{U}{L_z} \int\limits_{0}^{L_y} \mkern-3mu \deriv y \; \psi^y_i (y) \psi^y_f (y) \mkern-10mu \int\limits_{-L_z/2}^{+L_z/2} \mkern-12mu \deriv z \; e^{-i (k^z_i - k^z_f) z} \mkern-10mu \int\limits_{0}^{S(y,z)} \mkern-12mu \deriv x \; \psi^x_i (x) \psi^x_f (x),
\end{align*}
with barrier height $U$. The integral in the last line can be expanded in $S$ to obtain an expression linear in $S$, referred to as the first order approximation. Additionally, the limit of $U \rightarrow +\infty$ can be taken to obtain $U \psi^x_i (0) \psi^x_f (0) \rightarrow  2\sqrt{E^x_i E^x_f}/L_x$ with $E_{i/f}^x$ the energy of the initial and final state wave functions along the $x$-direction. This limit is known as the Prange-Nee approximation \cite{prange1968quantum}. If there are few subbands crossing the Fermi level, which is typically the case in semiconductors, the wave functions barely oscillate along the confinement and these commonly used approximations work well. In metallic nanowire simulations however, highly oscillating wave functions are present and these approximations lead to large estimation errors (see Fig.~\ref{WFInt} and Fig.~\ref{figScatProb}).

To obtain the scattering rate $P$, we need the absolute value squared of the matrix elements:
\begin{align*}
P\lef \mid i \rangle \rightarrow \mid f \rangle \rig &= \frac{2\pi}{\hbar} \left| \langle i \mid V^\textnormal{\tiny SR} \mid f \rangle \right|^2 \delta\lef E_i - E_f \rig.
\end{align*}
Because we are interested in an average scattering rate, determined by the SR standard deviation and correlation length, we have to average over this absolute value squared, making use of the statistics of the surface functions $S$. Because of the square, the averaging procedure needs a distribution function of $S$ at two different positions:
\begin{align}
\label{eqScatAv}
P^\textnormal{av.}\lef \mid i \rangle \rightarrow \mid f \rangle \rig \equiv \int \mkern-3mu \deriv S \int \mkern-3mu \deriv \tilde{S} \; f (S, \tilde{S}) P\lef \mid i \rangle \rightarrow \mid f \rangle \rig,
\end{align}
where the scattering rate $P$ depends on the two SR functions $S$ and $\tilde{S}$. It seems unlikely that an analytic expression can be obtained for the average scattering rate if no approximations of the matrix elements are made and the distribution function has to be convolved with a product of highly nonlinear expressions. Indeed, this appears to be the case for the bivariate normal distribution \cite{lizzit2014new}. However, an analytic solution was obtained by using distribution functions on a finite domain, not requiring numerical integration (very time consuming) to obtain the scattering rates and transport properties \cite{moors2015surfaceroughness}.

\subsection{Analytic solution with finite domain distribution functions}
We propose two variants of a distribution function on a finite domain $f_\textnormal{I/II}(S,\tilde{S})$ ($ -\sqrt{3}\Delta \leq S,\tilde{S}, \leq \sqrt{3}\Delta$) that capture the SR statistics with standard deviation $\Delta$, correlation length $\Lambda$ and Gaussian autocorrelation $\langle S \tilde{S} \rangle = \Delta^2 e^{-(\mathbf{r}-\tilde{\mathbf{r}})^2/(\Lambda^2/2)}$:
\begin{align*}
f_\textnormal{I} ( S, \tilde{S} ) &\equiv \frac{1}{12\Delta^2}\left\{ 1 + \frac{4}{3} C \left[ \theta ( S \tilde{S} ) - \theta ( - S \tilde{S} ) \right] \right\}, \\
f_\textnormal{II} ( S, \tilde{S} ) &\equiv \frac{1}{12\Delta^2}( 1 - C ) + \frac{1}{\sqrt{12} \Delta} C \delta ( S - \tilde{S} ), \\
C &\equiv e^{-(\mathbf{r}-\tilde{\mathbf{r}})^2/(\Lambda^2/2)},
\end{align*}
with $\theta$ the step function. Both distribution functions are less realistic (negative weights when $C>3/4$ for variant I and a distribution that peaks at the diagonal $S=\tilde{S}$ for variant II) than the bivariate normal distribution function, being the natural choice, but the results were shown to be in very good agreement in a one subband toy model \cite{moors2015surfaceroughness}. The integration in Eq.~(\ref{eqScatAv}) can be performed and this expression can be explicitly written in a function of a Boltzmann solver \cite{moors2015surfaceroughness}.

\section{Results \& Discussion}
\begin{figure}
\centering
\subfigure[\ PN: $D\approx 2.15$~nm]{\includegraphics[width=0.215\linewidth]{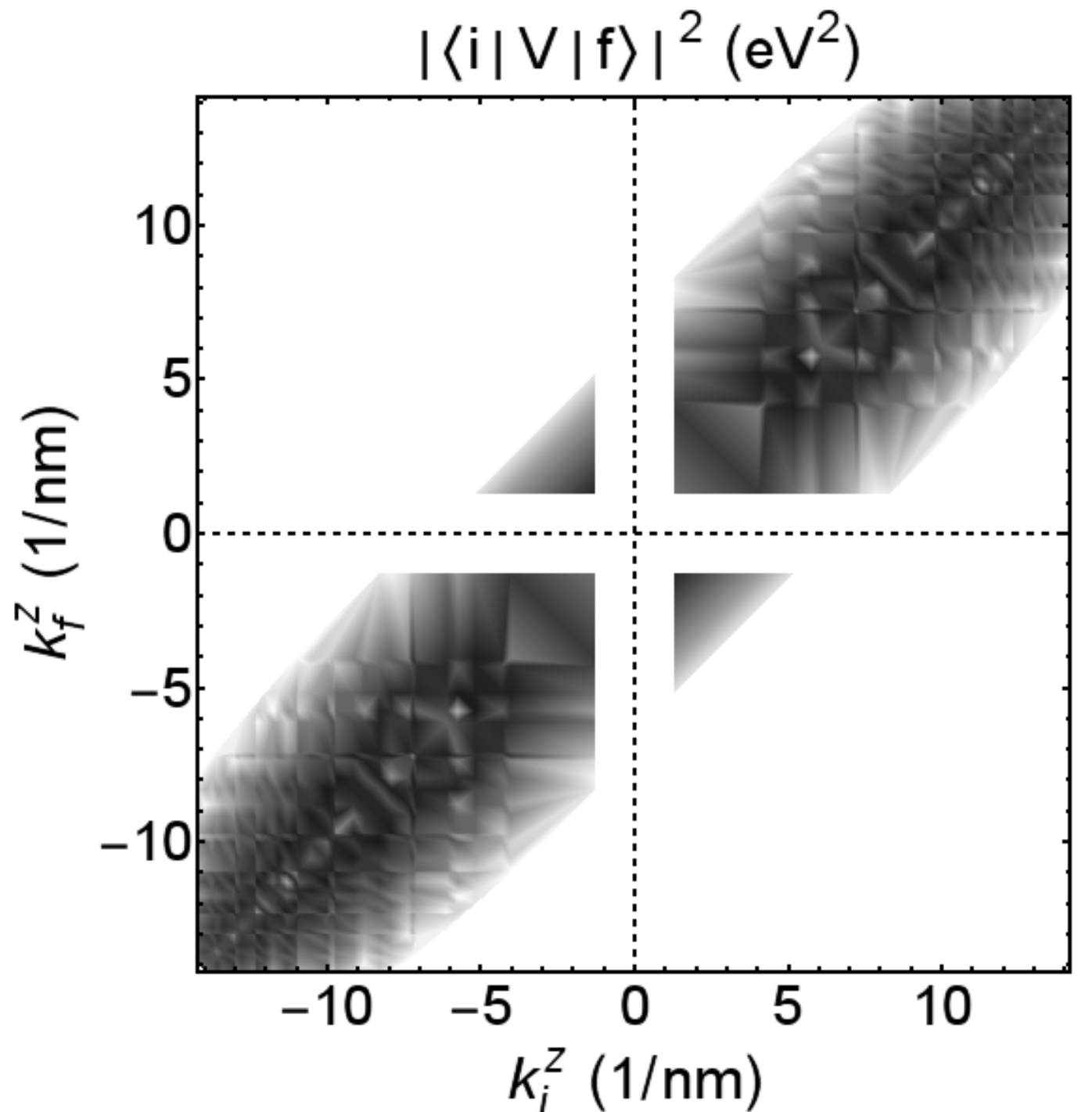}}
\subfigure[\ PN: $D\approx 3.25$~nm]{\includegraphics[width=0.26\linewidth]{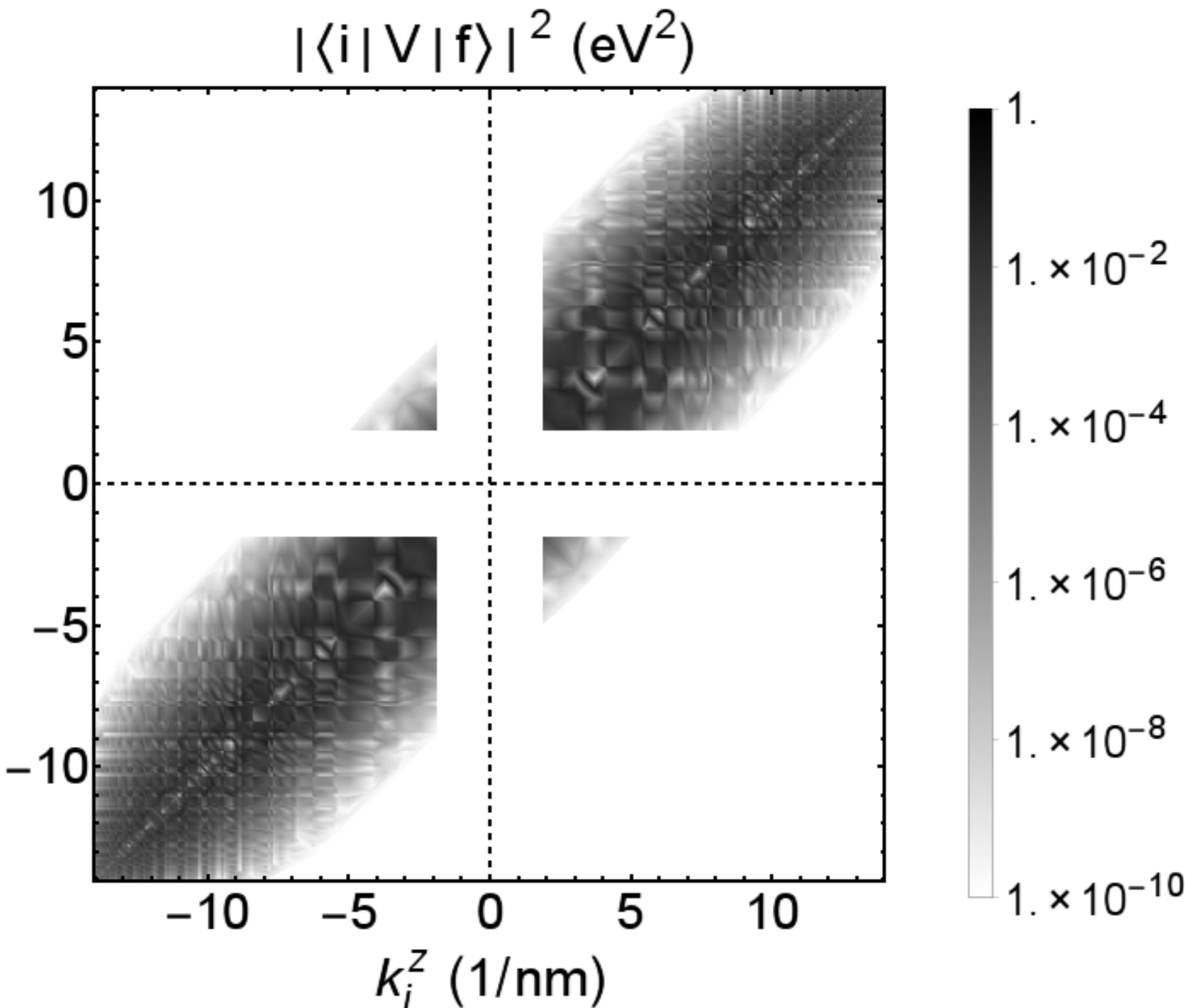}}
\subfigure[\ FO: $D\approx 2.15$~nm]{\includegraphics[width=0.215\linewidth]{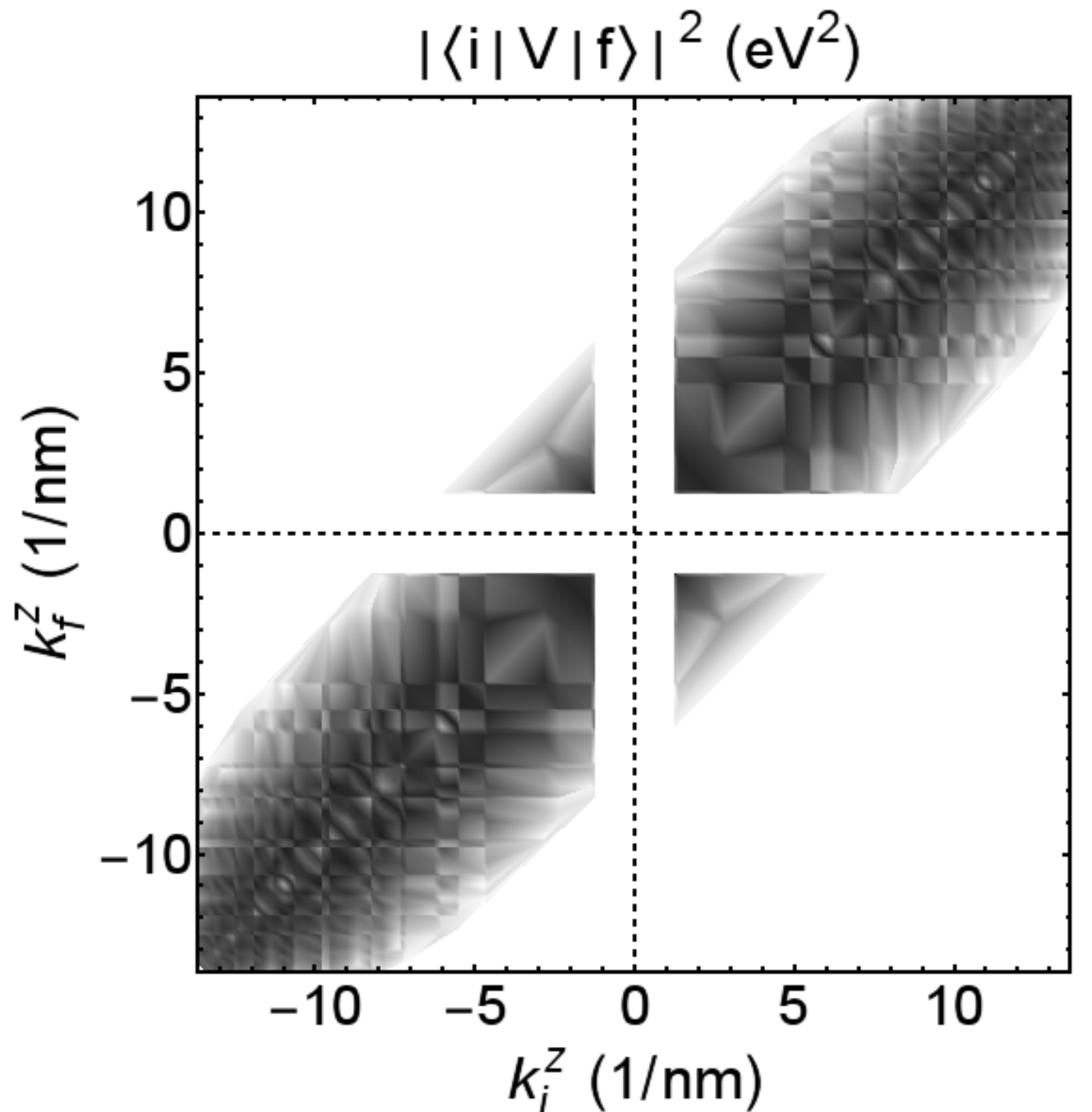}}
\subfigure[\ FO: $D\approx 3.25$~nm]{\includegraphics[width=0.26\linewidth]{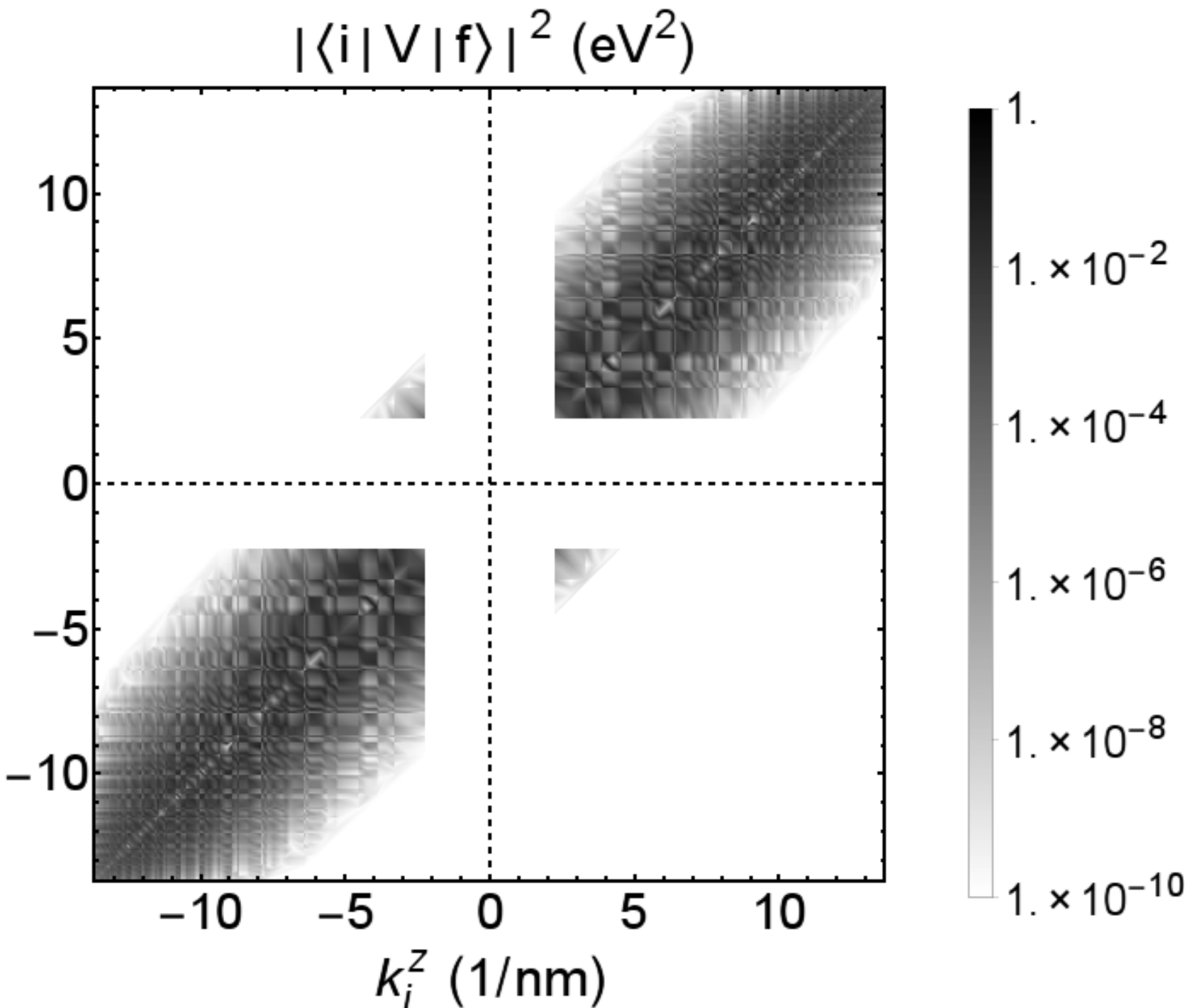}}
\subfigure[\ FD I: $D\approx 2.15$~nm]{\includegraphics[width=0.215\linewidth]{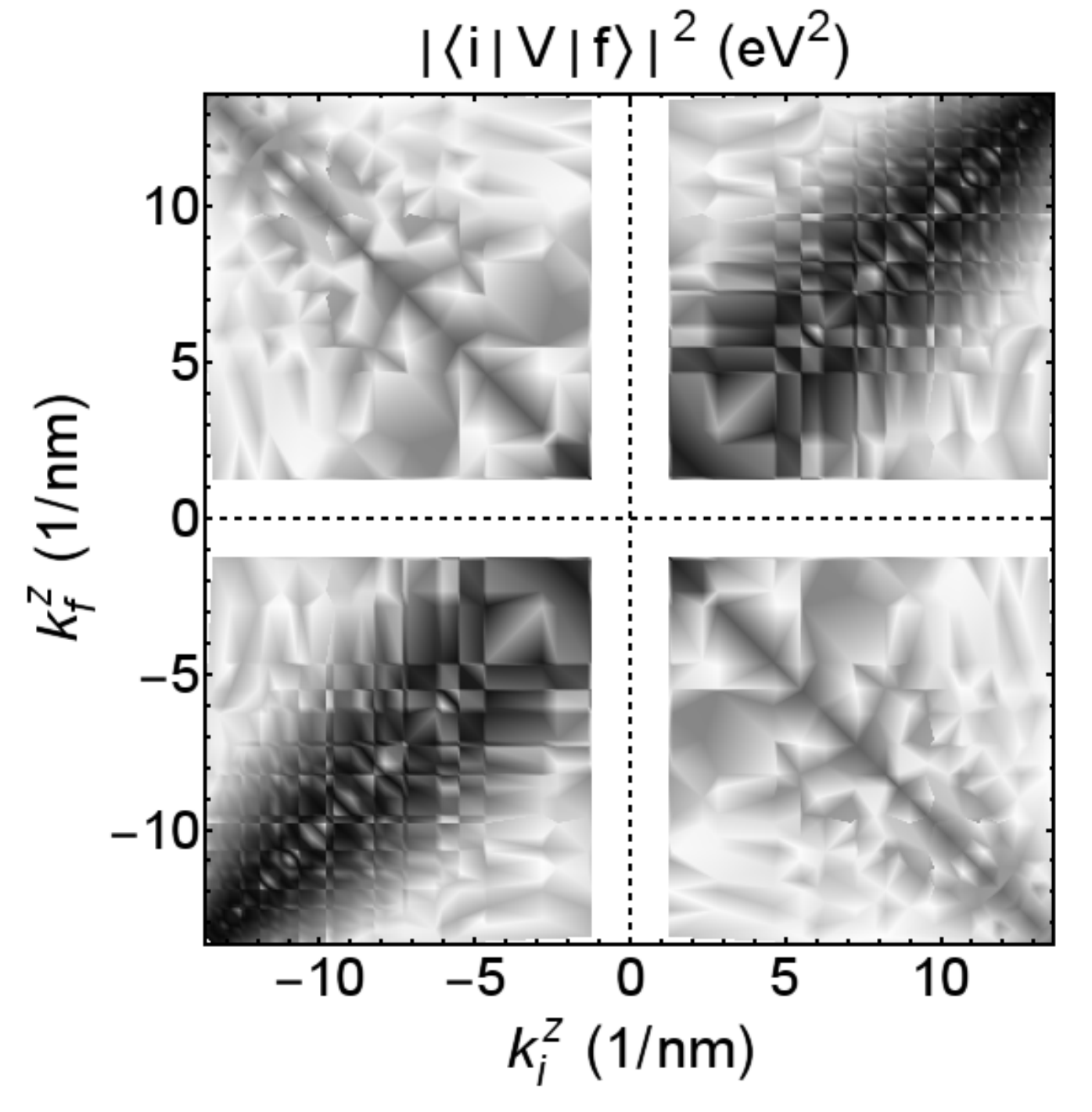}}
\subfigure[\ FD I: $D\approx 3.25$~nm]{\includegraphics[width=0.26\linewidth]{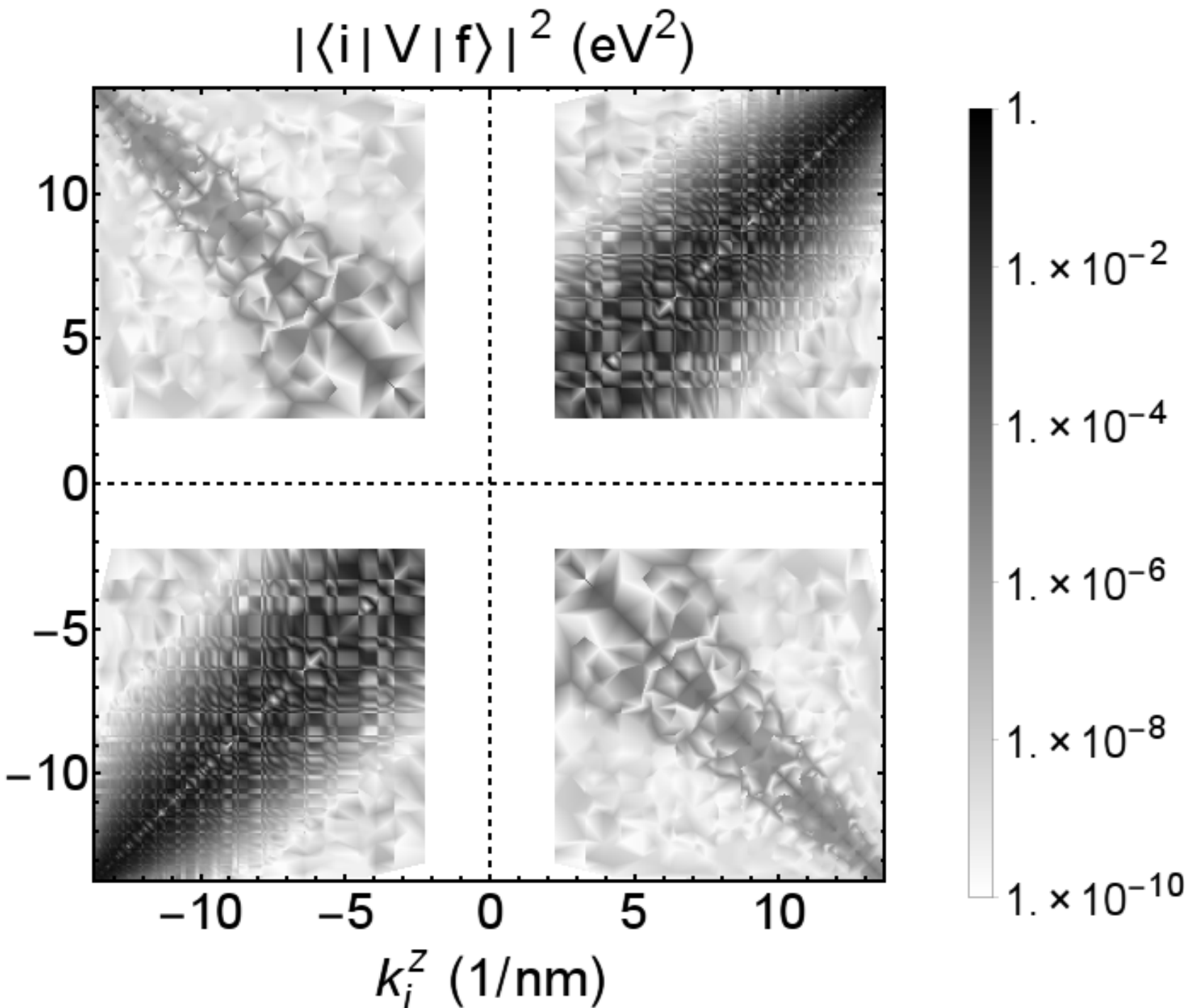}}
\subfigure[\ FD II: $D\approx 2.15$~nm]{\includegraphics[width=0.215\linewidth]{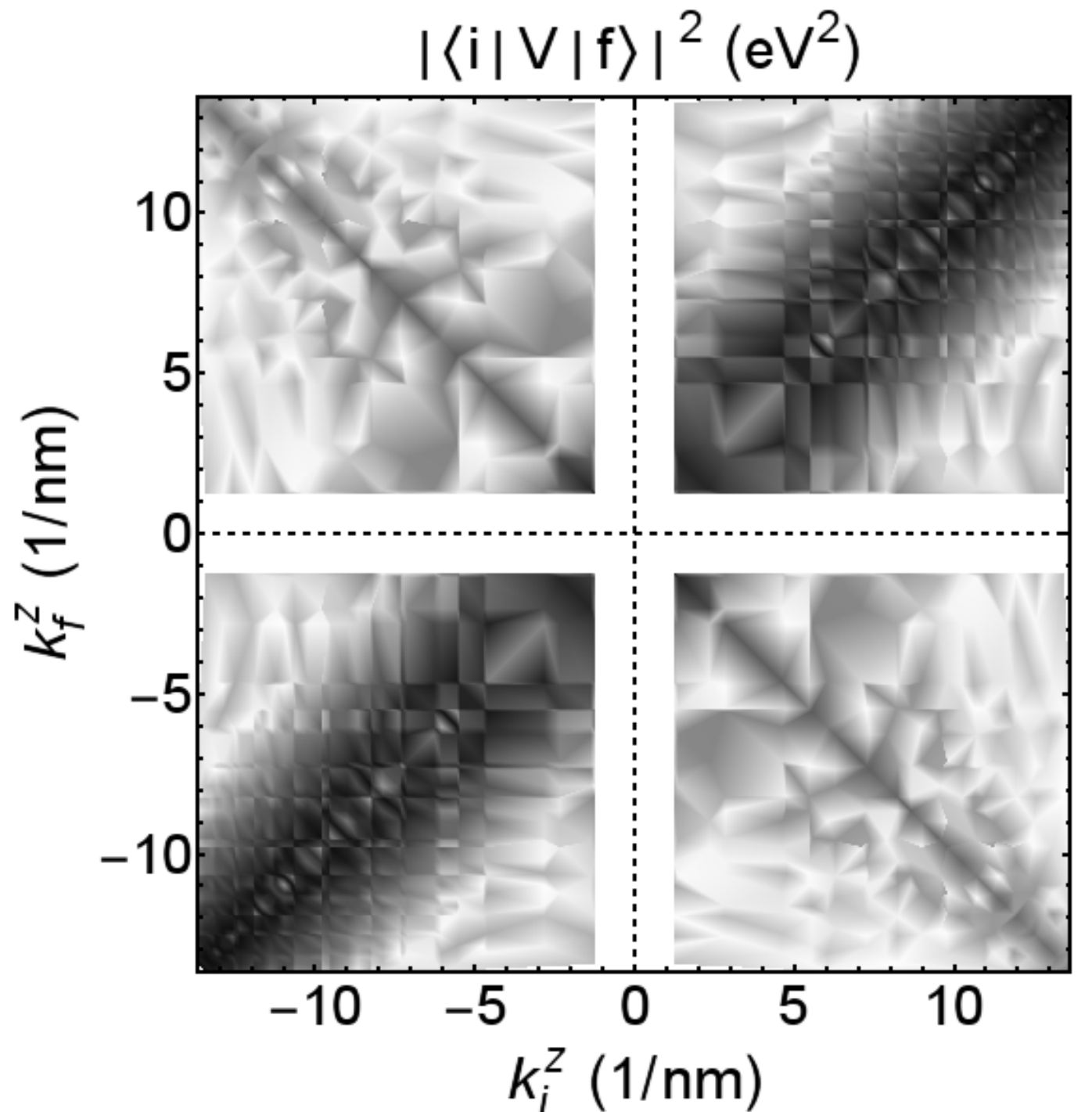}}
\subfigure[\ FD II: $D\approx 3.25$~nm]{\includegraphics[width=0.26\linewidth]{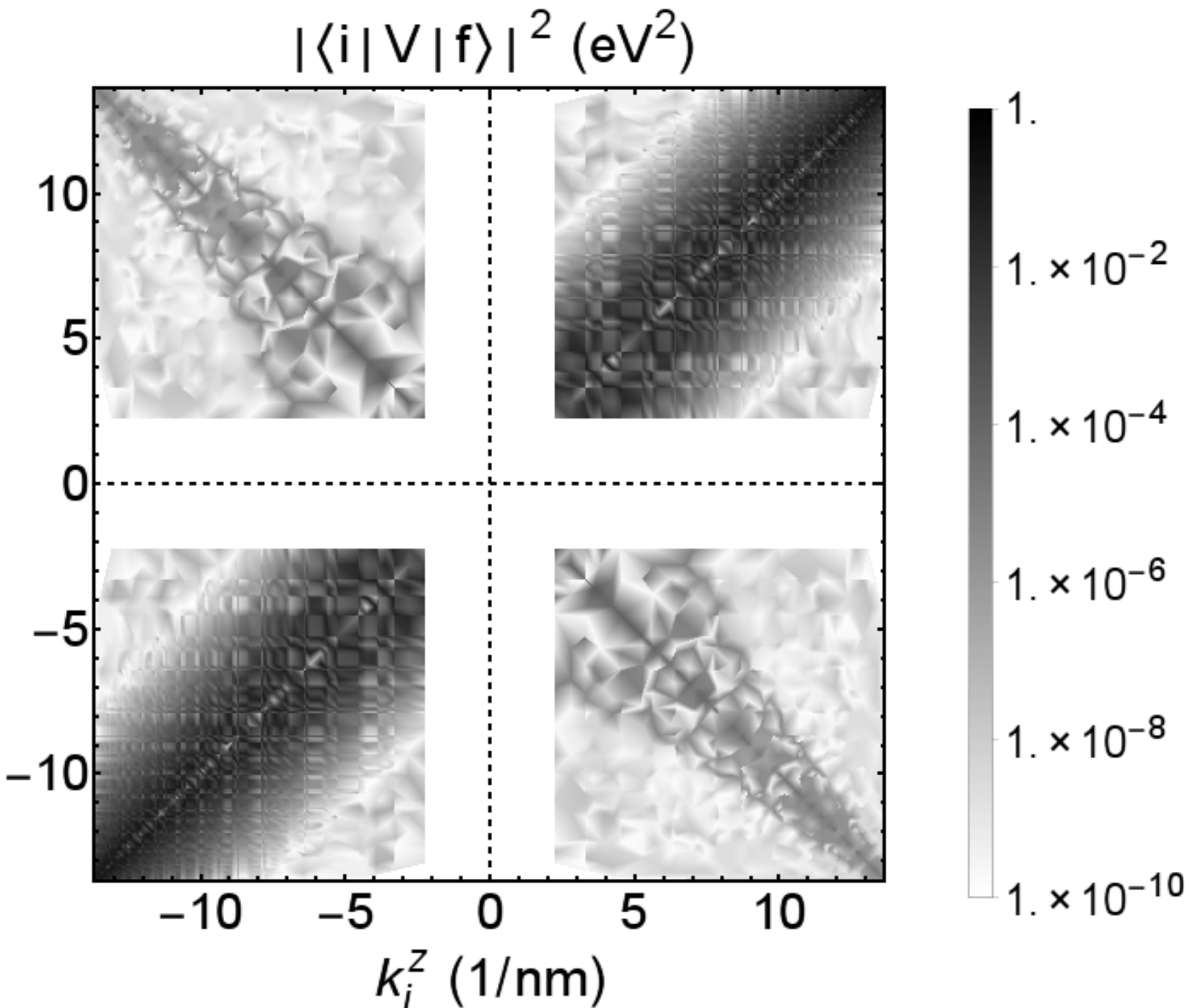}}
\caption{The values of the averaged SR matrix elements are shown in a logarithmic scale as a function of the wavevector along the transport direction of initial and final state. The SR matrix elements are obtained with the Prange-Nee (PN) and first order (FO) approximation and the two variants of the finite domain methods (FD I/II) for two Cu nanowires (the parameters are the same as in Fig.~\ref{figCond}) with different diameters. In case of degeneracy of state pairs with labels $(k_i^z, k_f^z)$, the maximal matrix element value is shown.}
\label{figScatProb}
\end{figure}

The averaged absolute value squared of the SR matrix elements, obtained with the approximated wavefunction overlap models (first order and Prange-Nee approximation) and finite domain models, are shown in Fig.~\ref{figScatProb} for two nanowires with different diameters. Scattering is mostly occurring between states that have a very similar wavevector while back-scattering is peaked around the Umklapp processes. The effect of subband quantization can also be clearly seen in the color plots, as well as the gap in wavevector space around $k^z=0$ because of a lack of subbands crossing the Fermi level there. The size of this gap is very different for the two diameters (see Fig.~\ref{figSubbands}) and this has a visible impact on the back-scattering components in the off-diagonal quadrants: a larger the gap reduces the back-scattering probabilities. There is also a large difference between the approximated and unapproximated wave function overlap methods, while the two variants of the finite domain models are in very good agreement.

\begin{figure}[tb]
\centering
\subfigure[\ Resistivity]{\includegraphics[width=0.425\linewidth]{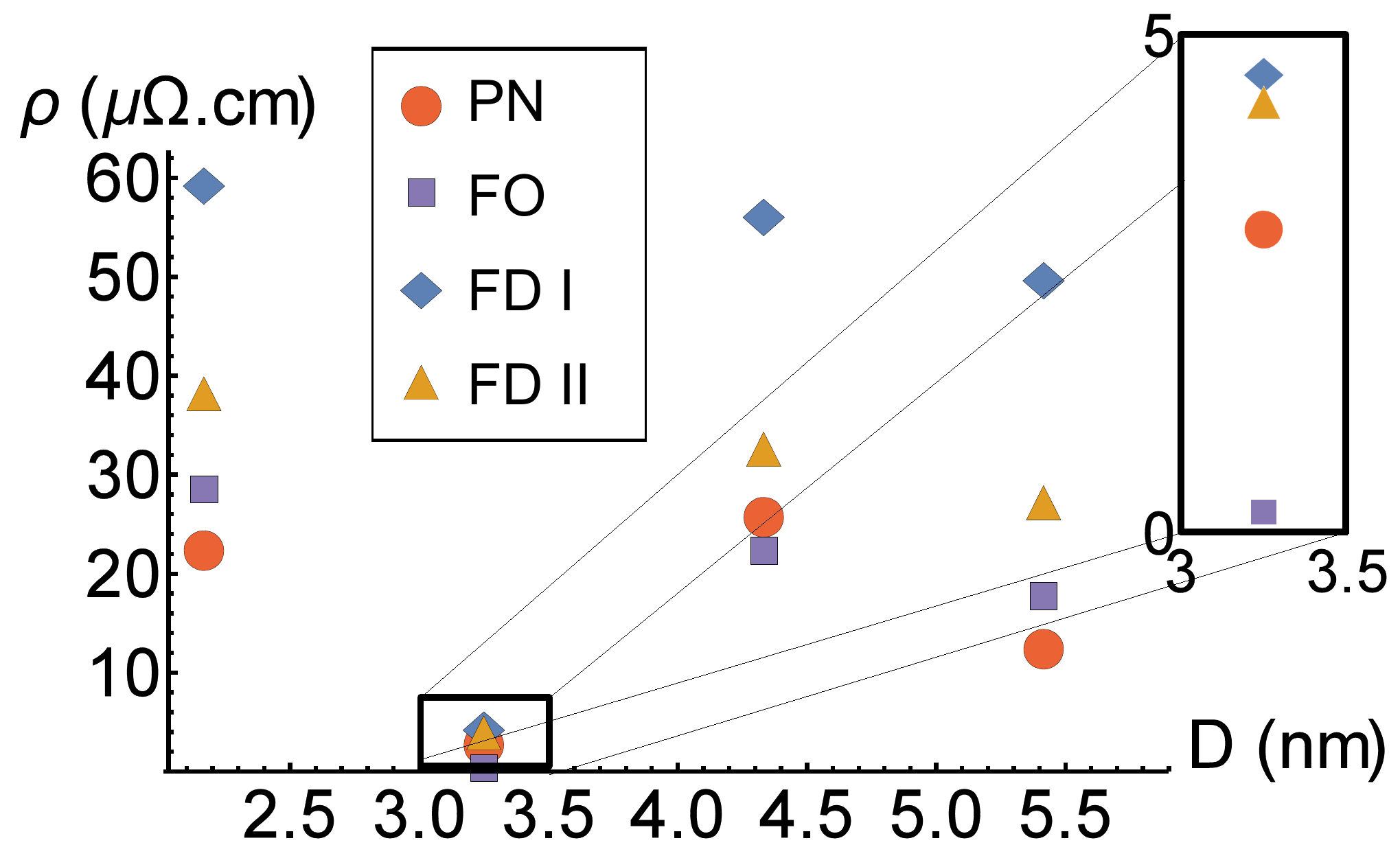}}
\subfigure[\ Wavevector gaps]{\includegraphics[width=0.45\linewidth]{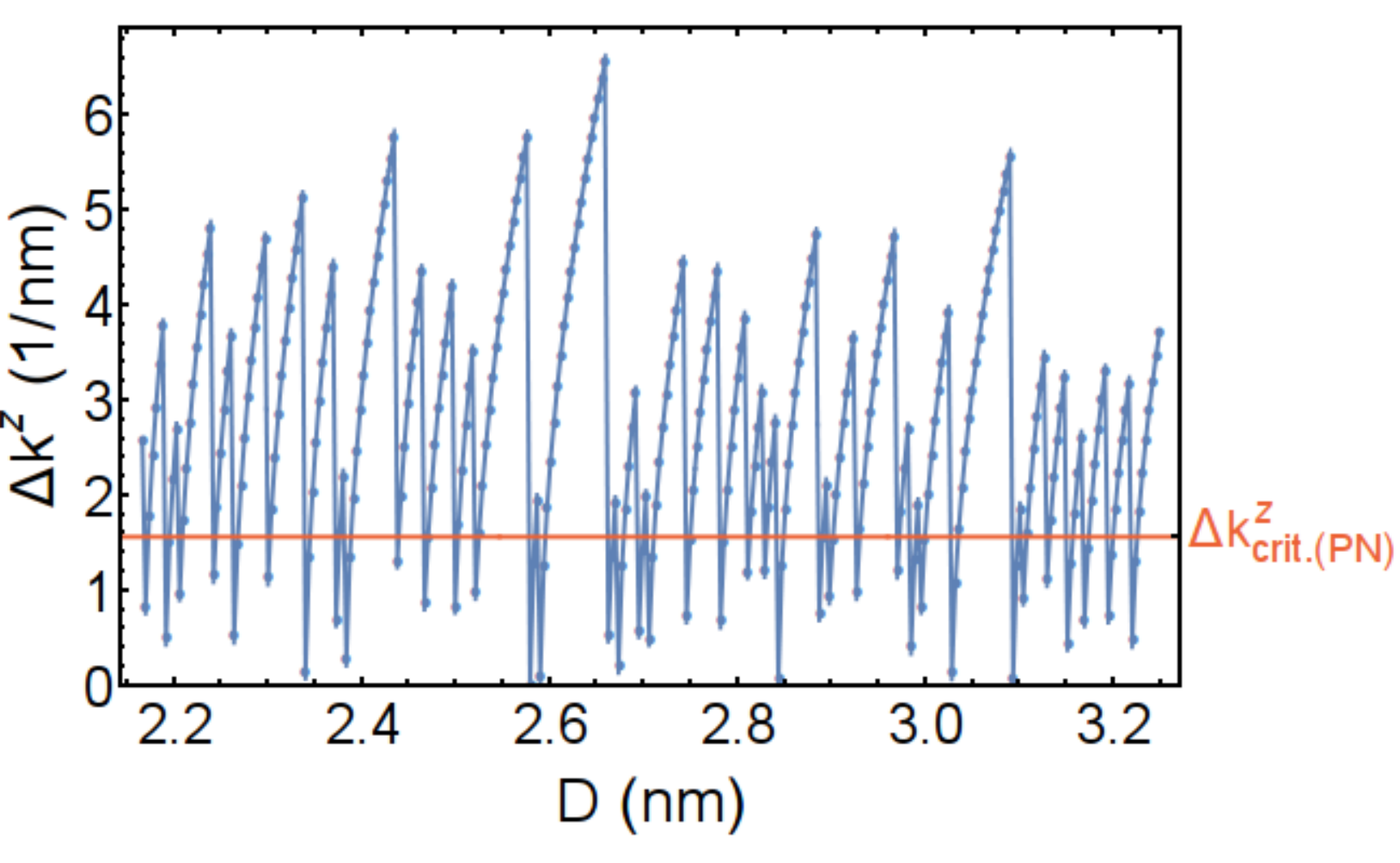}}
\caption{The resistivity for Cu nanowires with square cross section and sides $D$, ranging from 2 to 6~nm (SR standard deviation $\Delta = 2a_\textnormal{\tiny Cu} \approx 0.7$~nm and correlation length $\Lambda = 5a_\textnormal{\tiny Cu} \approx 1.75$~nm), are shown, making use of the two variants of the finite domain methods (FD I/II) and the first order (FO) and Prange-Nee (PN) approximation for the SR matrix elements. The resistivity will drop repeatedly in between two data points, always when a substantial wavevector gap (or correspondingly momentum gap) appears between left- and right-moving electrons at the Fermi level. The wavevector gap and the critical gap $\Delta k^z_\textnormal{crit. (PN)} \equiv 2 \sqrt{2} / \Lambda$, obtained with Prange-Nee, are shown in (b) for diameters in between $D\approx 2.15$~nm and $D\approx 3.25$~nm \cite{moors2015surfaceroughness}.}
\label{figCond}
\end{figure}

The resistivity obtained with a self-consistent multi-subband Boltzmann solver, using the first order and Prange-Nee approximation and the two variants of the finite domain models, is shown in Fig.~\ref{figCond} (a) for four different diameters. All methods predict a resistivity of the same order of magnitude, the scaling being very different from the larger diameter 1/width proportionality. The $D\approx 3.25$~nm wire shows a very low resistivity compared to the other values, with good agreement for the finite domain models but substantial deviations for the Prange-Nee and first order approximations. A low resistivity appears to be related to the presence of a large wavevector gap $\Delta k^z$ between the left- and right-moving electron states crossing the Fermi level (see Fig.~\ref{figSubbands} and Fig.~\ref{figCond} (b)). It was shown that the resistivity is exponentially dependent on the SR correlation length; hence short length scale SR induces quite more current loss than long length scale SR. The resistivity also drops exponentially with $\Delta k^z$ when $\Delta k^z$ exceeds a certain critical value \cite{moors2015surfaceroughness}. With the Prange-Nee approximation this critical gap, as a function of correlation length $\Lambda$ can be estimated as $\Delta k^z_\textnormal{crit. (PN)} \equiv 2\sqrt{2}/\Lambda$. Looking at the wavevector gaps as a function of the diameter in Fig.~\ref{figCond} (b), a nanowire protected from back-scattering can only be realized when the average diameter is controlled up to an angstrom-scale resolution. This diameter window can be enlarged by using metallic nanowires with a lower conduction electron density and fewer subbands crossing the Fermi level.

\section{Conclusion}
We have introduced two variants of distribution functions on a finite domain to obtain analytic solutions of an average scattering rate for surface roughness scattering using Ando's model. It allows for fast and accurate simulations of metallic nanowires to study the impact of surface roughness on the transport properties. The analytic solutions prevent us from having to resort to approximations of the matrix elements by expanding the wavefunction overlap for small roughness sizes or even taking the infinite barrier limit, completely neglecting the oscillations of the wavefunctions normal to the wire boundary surface. This appears to be important in metallic nanowires where many subbands cross the Fermi level with highly oscillating wavefunctions in the confinement directions. When there is a substantial wavevector gap between left- and right-moving electrons at the Fermi level, back-scattering is suppressed and the forward current is protected, resulting in a very low resistivity. The predicted resistivity values are in good agreement for the finite domain models and often also for the approximated models, but they poorly estimate the wave function overlap; hence this agreement should rather be viewed as a coincidence.

\bibliographystyle{IEEEtran}
\bibliography{IEEEabrv,sample}

\end{document}